\documentstyle[aps,preprint,pra,epsfig]{revtex} 

\begin{document} 
 
\draft 
 
\title{Observable Signature of the Berezinskii-Kosterlitz-Thouless Transition  
in a Planar Lattice of Bose-Einstein Condensates} 
 
\author{A. Trombettoni$^{1}$, A. Smerzi$^{2,3}$, and P. Sodano$^{4}$} 
\address{$^1$ Istituto Nazionale per la Fisica della Materia and 
Dipartimento di Fisica, Universita' di Parma, 
parco Area delle Scienze 7A, I-43100 Parma, Italy\\ 
$^2$ Istituto Nazionale per la Fisica della Materia BEC-CRS and 
Dipartimento di Fisica, Universita' di Trento, I-38050 Povo, Italy\\ 
$^3$ Theoretical Division, Los Alamos 
National Laboratory, Los Alamos, NM 87545, USA\\ 
$^4$ Dipartimento di Fisica and Sezione I.N.F.N., Universit\`a di  
Perugia, Via A. Pascoli, I-06123 Perugia, Italy\\ 
} 
\date{\today} 
\maketitle 
 
\begin{abstract} 
We investigate the possibility that Bose-Einstein condensates 
(BECs), loaded on a $2D$ optical lattice, undergo - at finite temperature -  
a Berezinskii-Kosterlitz-Thouless (BKT) transition.  
We show that - in an experimentally attainable range  
of parameters - a planar lattice of BECs is described by the $XY$ model  
at finite temperature. We demonstrate that  
the interference pattern of the expanding condensates provides  
the experimental signature of the BKT transition  
by showing that, near the critical temperature,  
the $\vec{k}=0$ component of the momentum distribution and  
the central peak of the atomic density profile sharply decrease. 
The finite-temperature transition for a $3D$ optical lattice 
is also discussed, and the analogies with superconducting 
Josephson junction networks are stressed through the text. 
\end{abstract} 
 
 
\section{Introduction} 
 
Recent advances in the manipulation of cold atomic gases  
allow nowadays for storing Bose-Einstein condensates (BECs) in $1D$  
\cite{anderson98,cataliotti01,morsch01,hensinger01,eiermann03},  
$2D$ \cite{greiner01} and $3D$ \cite{greiner02} optical lattices,  
opening a pathway for the experimental and theoretical analysis of finite  
temperature transitions in atomic systems. 
In a way similar to the spectacular  
realizations obtained with superconducting Josephson networks  
\cite{beasley79,resnick81}, arrays of BECs can provide 
new experimentally realizable systems on which to test well known   
paradigms of the statistical mechanics: our interest here 
is to evidence how to detect a finite-temperature defect-mediated  
transition in $2D$ optical lattice of BECs.  
 
According to the well-known Mermin-Wagner theorem \cite{nelson83},  
two-dimensional systems with a continuous symmetry 
cannot sustain long-range order in the thermodynamic limit  
at finite temperature; however,  
a phase transition is still possible and it occurs via the 
unbinding of point defects like dislocations or vortices.  
Defect-mediated transitions are widely studied \cite{nelson83} since  
they provide crucial insights for a wide variety of 
experiments on thin films. The BKT transition is the 
paradigmatic example of a defect-mediated transition \cite{nelson83}  
which is exhibited by the two-dimensional  
$XY$ model \cite{minnaghen87,kadanoff00},  
describing $2$-components spins on a two-dimensional lattice.  
In the low-temperature phase, characterized by the presence of  
bound vortex-antivortex pairs, the spatial correlations exhibit a power-law  
decay; above a critical temperature $T_{BKT}$, the decay is exponential  
and there is a proliferation of free vortices.  
The BKT transition has been observed in superconducting Josephson arrays  
\cite{resnick81} and its predictions are well verified in  
the measurements of the superfluid density  
in $^4He$ films \cite{bishop78}. 

In finite magnetic systems the BKT transition point  
is signaled by the drop to $0$ of a suitably defined 
magnetization \cite{berezinskii73}; the existence  
of such a magnetization in finite systems does not contradict  
the Mermin-Wagner theorem, since the latter 
is valid only in the thermodynamic limit.  
In a large class of experimental situations, the finite size  
$XY$ model predicts magnetization exponents 
in agreement with the measurements  
carried out for layered magnets with planar spin symmetry \cite{bramwell93}. 
We shall show in the sequel that the Fourier transform 
at $\vec{k}=0$ of the condensates wavefunction provide 
the analogous of this magnetization for atomic systems, 
which are naturally finite (since they are trapped by an harmonic potential).
 
In this paper we show that - at a critical temperature $T_{BKT}$ 
lower than the temperature $T_{BEC}$ at which condensation  
in each well occurs - $2D$ lattices of BECs may undergo a 
phase transition to a superfluid regime  
where the phases of the single-well condensates are coherently aligned 
allowing for the observation  
of a Berezinskii-Kosterlitz-Thouless (BKT) transition.  
We show in the following that the recently realized  
$2D$ optical lattice of Bose-Einstein  
condensates \cite{greiner01} may allow for the observation  
of a BKT transition in a finite bosonic system, 
since the thermodynamic properties  
of the bosonic lattice at finite temperature may be well described,  
in a suitable and experimentally attainable range of parameters,  
by the Hamiltonian of the $XY$ model. It is easy to convince oneself 
that the transition we propose to observe  
is different from the quantum phase transition reported in \cite{greiner02}, 
where the system is at $T=0$ and the insulator phase 
(signaled by a reversible 
destruction of the phase coherence across the lattice) 
is reached varying the optical lattice  
parameters: at variance, here, 
we propose to fix the system parameters in the superfluid phase and  
increase the temperature $T$ until the thermally induced vortex  
proliferation determines the BKT transition.   
 
We shall also show that the experimental signature  
of the BKT transition in bosonic planar lattices  
is obtained by measuring, as a function of the temperature,  
the central peak of the interference pattern  
of the expanding condensates released  
from the trap when the confining potentials are switched off.  
In fact, the peak of the momentum distribution at  
$\vec{k}=0$ may be regarded as the magnetization of a  
finite size $2D$ $XY$ magnet, and,  
as shown in the following, it should exhibit a sharp decrease  
around $T_{BKT}$. As a consequence, also the central peak of the atomic  
density profile decreases around $T_{BKT}$, signaling the occurrence  
of the BKT transition. 

The plan of the paper is the following: 
in Section II we discuss the Hamiltonian  
describing a $2D$ lattice of Bose-Einstein condensates at finite temperature  
and we show that - in a suitable range of parameters - 
the system is described by 
the $XY$ Hamiltonian. 
In Section III we show how the signature of the BKT transition 
may be evidenced in a bosonic planar lattice; there we shall also 
discuss interesting analogies 
with superconducting Josephson junction networks.
Section IV is devoted to our concluding remarks.
In Appendix A we report the details of a variational estimate 
of the coefficients of the Hamiltonian describing 
$1D$, $2D$ and $3D$ optical lattices, while Appendix B contains the  
analytical computation of the Fourier transform of a vortex on a lattice.  
 
\section{The Finite-Temperature Hamiltonian for the System} 
 
For $2D$ optical lattices, when the polarization vectors of the two  
standing wave laser fields are orthogonal, the resulting periodic  
potential for the atoms is 
\begin{equation} 
V_{opt}(\vec{r})=V_0 [ \sin^2{(kx)} + \sin^2{(ky)} ] 
\label{opt_potential_2D} 
\end{equation} 
where $k=2 \pi/\lambda$ is the wavevector of the laser beams.  
The potential maximum of a single standing wave,   
$V_0=sE_R$, may be controlled by changing the intensity of the optical  
lattice beams, and it is conveniently measured in units of the recoil energy 
$E_R=\hbar^2 k^2/2m$ ($m$ is the atomic mass),  
while, typically, $s$ can vary from $0$ up to $30$ (gravity  
is assumed to act along the $z$-axis).  

Usually, superimposed to the optical potential, 
there is an harmonic magnetic potential   
\begin{equation} 
V_{m}(\vec{r})=\frac{m}{2} [ \omega_r^2 (x^2+y^2) + \omega_z^2 z^2 ], 
\label{magn_potential_2D} 
\end{equation} 
where $\omega_z$ ($\omega_r$) is the axial (radial) trapping frequency.  
The minima of the $2D$ periodic potential (\ref{opt_potential_2D}) 
are located at the points $\vec{j}=(j_1,j_2) \cdot \frac{\lambda}{2}$  
with $j_1$ and $j_2$ integers, and the potential around the minima is 
$V_{opt} \approx \frac{m}{2} \tilde{\omega}_r^2 [ (x-\lambda j_1/2)^2 +   
(y-\lambda j_2/2)^2 ]$ with  
\begin{equation} 
\tilde{\omega}_r=\sqrt{2 V_0 k^2 / m} 
\label{omega_tilde} 
\end{equation} 
providing the frequency of the wells. 
When $\tilde{\omega}_r \gg \omega_r,\omega_z$,  
the system realizes a square array of tubes, i.e. an array of harmonic traps  
elongated along the $z$-axis \cite{greiner01}.  
Although the axial dynamics is very interesting in its  
own right \cite{pedri03}, in the following we analyze only  
the situation in which the axial degrees of freedom are frozen out,  
which is realizable if $\omega_z$ is sufficiently large. We also 
assume that the harmonic oscillator length  
of the magnetic potential $\sqrt{\hbar/m \omega_r}$ is  
larger than the size $L$ (in lattice units) of the optical  
lattice in order to reduce density inhomogeneity effects and to allow for  
a safe control of finite size effects. 
In addition, when all the relevant energy scales  
are small compared to the excitation  
energies one can safely expand the field operator \cite{jaksch98} as  
\begin{equation} 
\hat{\Psi}(\vec{r},t)= \sum_{\vec{j}} \hat{\psi}_{\vec{j}}(t)  
\Phi_{\vec{j}}(\vec{r}) 
\label{TB}  
\end{equation} 
with $\Phi_{\vec{j}}(\vec{r})$ 
the Wannier wavefunction localized at the $\vec{j}$-th well  
(normalized to $1$) and $\hat{N}_{\vec{j}}=\hat{\psi}^{\dag}_{\vec{j}}  
\hat{\psi}_{\vec{j}}$ the bosonic number operator. 
Substituting the expansion in the full quantum Hamiltonian  
describing the bosonic system, leads to  
the Bose-Hubbard model \cite{fisher89,jaksch98}  
\begin{equation} 
\hat{H}=-K \sum_{<\vec{i}, \vec{j}>}  
(\hat{\psi}^{\dag}_{\vec{i}} \hat{\psi}_{\vec{j}}+ h.c.) 
+ {U \over 2} \sum_{\vec{j}} \hat{N}_{\vec{j}}  
(\hat{N}_{\vec{j}}-1)  
\label{BHM} 
\end{equation} 
where $\sum_{<\vec{i}, \vec{j}>}$ denotes a sum over  
nearest neighbours, $ U = (4 \pi \hbar^2 a / m) \int d\vec{r} \,  
\Phi_{\vec{j}}^4$ ($a$ is the $s$-wave scattering length) and  
$K \simeq - \int d\vec{r} \, \big[ \frac{\hbar^2}{2m}  
\vec{\nabla} \Phi_{\vec{i}} \cdot \vec{\nabla} \Phi_{{\vec{j}}} +  
\Phi_{\vec{i}}  V_{ext} \Phi_{\vec{j}} \big]$.  
  
Upon defining  
\begin{equation} 
J = 2 K N_0 
\label{Jos_energy} 
\end{equation}  
(where $N_0$ is the average number of atoms per site),    
when $N_0 \gg 1 $ and $J/N_0^2 \ll U$, the Bose-Hubbard model reduces to  
\begin{equation} 
\hat{H}=H_{XY}- {U \over 2}\sum_{\vec{j}}  
\frac{\partial^2}{\partial \theta_{\vec{j}}^2}, 
\label{QPM} 
\end{equation}  
which describes the so-called quantum phase model  
(see e.g. \cite{simanek94,fazio01}): in Eq.(\ref{QPM}) 
$\theta_{\vec{j}}$ is the phase of the $j$-th condensates and  
\begin{equation} 
H_{XY}=-J \sum_{<\vec{i},\vec{j}>} \cos{(\theta_{\vec{i}}- 
\theta_{\vec{j}})} 
\label{X-Y} 
\end{equation} 
stands for the Hamiltonian of the classical $XY$ model.  
When $J \gg U$ and at temperatures $T \gg U/k_B$, 
the pertinent partition function  
describing the thermodynamic behaviour  
of the BECs stored in an optical lattice may be computed  
using the classical $XY$ model (\ref{X-Y}). 
 
The close relationship of the Bose-Hubbard model (\ref{BHM}) and 
the quantum phase model (\ref{QPM}) is by now well known 
\cite{fisher88} and it has been  
widely used in the analysis of superconducting Josephson networks; 
since, in superconducting networks, 
$N_0$ is the average number of excess Cooper pairs at a given site, 
the requirement $N_0 \gg 1 $ is   
of course satisfied and the condition $J/N_0^2 \ll U$ may 
also be easily realized 
for reasonable values of $J$. 
At variance, in bosonic lattices  
$N_0$ varies usually between $\sim 1$ and $\sim 10^3$ and the validity  
of the mapping between the Bose-Hubbard model and 
the quantum phase Hamiltonian is not always guaranteed; 
furthermore, in order to get the $XY$ model (\ref{X-Y}) from the  
Bose-Hubbard model (\ref{BHM}) $U$ should be much smaller than $J$, 
but not vanishing,  
in order to satisfy to the condition $J/N_0^2 \ll U$.  
Of course the Bose-Hubbard model - for $U=0$ -   
describes harmonic oscillators and  
thus cannot sustain any BKT transition; nevertheless, 
when $N_0 \gg 1 $ and $J/N_0^2 \ll U \ll J$, 
the Bose-Hubbard Hamiltonian reduces in the $XY$ model  
(\ref{X-Y}), which {\em do} display the BKT transition.  
 
A simple estimate of the coefficients $J$ and  
$U$ may be obtained by  
approximating the Wannier functions in Eq.(\ref{TB}) 
with gaussians, whose widths are 
determined variationally. 
The details of the computation are reported in Appendix A.  
In Fig.1 we plot $J/U$ as a function of $V_0$  
for a $2D$ optical lattice; for comparison, we plot the same ratio  
for $1D$ and $3D$ optical lattices also.  
Since, in a mean-field approach, the Mott insulator-superfluid transition  
occurs at $T=0$ approximately at $2 z J/U \sim 1$ \cite{fisher89,fazio01}  
($z$ is the number of nearest-neighbours), from Fig.1  
one sees that it is much easier to detect a quantum phase transition   
for the three-dimensional array (and indeed it has been recently detected   
\cite{greiner02}); no quantum phase transition has yet been  
reported in literature for $2D$ and $1D$ optical lattices, since   
a larger laser power $V_0$ is required for its observation.  
 
For a $2D$ lattice with $V_0$ between $20$ and $25E_R$ (and  
$N_0 \approx 170$ as in \cite{greiner01}),  
the conditions $J \gg U \gg J/N_0^2$ are rather well  
satisfied and the BKT critical temperature, $T_{BKT} \sim J/k_B$,  
is between $10$ and $30nK$. Since the BKT transition for the $XY$ model  
occurs at a temperature $T_{BKT} \sim J/k_B$, 
it should be possible to evidence  
the transition with measurements performed  
at different temperatures. 
Using a coarse-graining approach to determine the finite-temperature  
phase-boundary line of the Bose-Hubbard model \cite{kampf93}, 
yields, in $2D$ and for $N_0 \gg 1$ and $J / U \gg 1 $, 
a critical temperature $T_{BKT} \approx J / k_B$, 
in reasonable agreement with the $XY$ estimates of $T_{BKT}$; 
in addition, in the chosen range of parameters, the condition  
$U N_0^2/J=U N_0/K \gg 1$, satisfies the finite-temperature stability 
criterion recently derived by Tsuchiya and Griffin  \cite{tsuchiya03}.  
 
We notice that the $XY$ Hamiltonian,  
in the limits $N_0 \gg 1$ and $J \gg U \gg J/N_0^2$, can be 
also retrieved   
extending the Gross-Pitaevskii Hamiltonian to  
finite temperature $T \gg U/k_B$.  
Replacing the tight-binding ansatz  
$\Psi(\vec{r},t)= \sum_{\vec{j}} \psi_{\vec{j}}(t)  
\Phi_{\vec{j}}(\vec{r})$ (where $\psi_{\vec{j}}$ are classical fields  
and not operators) in the Gross-Pitaevskii Hamiltonian  
one gets the lattice Hamiltonian  
\begin{equation}
H=-K \sum_{<\vec{i}, \vec{j}>}  
(\psi^{\ast}_{\vec{i}} \psi_{\vec{j}}+ c.c.) 
+ \frac{U}{2} \sum_{\vec{j}} N_{\vec{j}} (N_{\vec{j}}-1),
\label{CBHM}
\end{equation}  
which is the classical version of the Bose-Hubbard model   
($N_{\vec{j}} \equiv \mid \psi_{\vec{j}}\mid^2$).  
Writing $N_{\vec{j}}=N_0 + \delta N_{\vec{j}}$, one may  
neglect  
the quadratic terms [i.e. $(\delta N_{\vec{j}})^2$]  
in the hopping part of the Hamiltonian (\ref{CBHM}), which then
reduces in the considered limit to $H_{XY}$ (\ref{X-Y}).  
 
Accurate Monte Carlo simulations yield - for the  
$XY$ model - the BKT critical temperature  
$T_{BKT}=0.898J/k_B$ \cite{gupta88}.
When $U \ll J$, a BKT transition still occurs at a slightly  
lower critical temperature $T_{BKT}(U)$.  
Intuitively speaking, when $U$ increases, the superfluid region  
in the phase diagram decreases and, thus, one has $T_{BKT}(U) < T_{BKT}$  
\cite{fazio01}.  
An estimate of the dependence of $T_{BKT}(U)$ on $U$ obtained 
using a renormalization group approach has been reported in \cite{smerzi04}:  
when $U \ne 0$, but still $U/J \ll 1$, the only effect of the interacting  
term amounts to renormalize $J$ and to 
lower the BKT critical temperature \cite{rojas96}.  
For this reason in the following we shall present results from Monte Carlo  
simulations of the classical $XY$ model only.  

In numerical simulations of the finite $XY$ lattice one defines the critical  
temperature as the temperature $T_C(L)$  
at which the correlation length equals the size $L$ of the square lattice  
\cite{bramwell93}; of course, as $L \to \infty$, $T_C(L) \to T_{BKT}$.  
For $T=T_C(L)$, the BKT transition is signaled in a finite system  
by the fact that a suitably defined magnetization,  
defined by Eq.(\ref{M}), drops to zero \cite{bramwell93}.  
 
The emerging physical picture is the following:  
There are two relevant temperatures for the system,  
the temperature $T_{BEC}$, at which in each well there is a condensate,  
and the temperature $T_{BKT}$ at which the condensates phases  
start to coherently point in the same direction. Of course,  
in order to have well defined condensates phases one should have  
$T_{BKT} < T_{BEC}$. 
The critical temperature $T_{BEC}$ is given by $T_{BEC} \approx 0.94 \hbar N_0^{1/3}  
\bar{\omega}/k_B$ where $\bar{\omega}= (\tilde{\omega}_r^2 \omega_z)^{1/3}$  
\cite{pitaevskii03}. With the numerical values given in \cite{greiner01},  
$T_{BEC}$ turns out to be   
$ \gtrsim \ 500 nK$ for $s \gtrsim 20$. When $T < T_{BEC}$,  
the atoms in the well $\vec{j}$  
of the $2D$ optical lattice may be described by a macroscopic  
wavefunction $\psi_{\vec{j}}$.  
Furthermore, when the fluctuations around  
the average number of atoms per site $N_0 \gg 1$ are strongly suppressed,  
one may put, apart from the factor $\sqrt{N_0}$ constant across the array,   
$\psi_{\vec{j}}=e^{i \theta_{\vec{j}}}$.  
The temperature $T_{BKT}$ is of order  
of $J/k_B$: with the experimental parameters of  
\cite{greiner01} and $V_0$ between $20$ and $25E_R$,  
one has that $T_{BKT}$ is between $10$ and $30nK$, which is  
sensibly smaller than the condensation temperature  
$T_{BEC}$ of the single well.  

A similar scenario describes   
also the phases of planar arrays of superconducting Josephson junctions  
\cite{simanek94,fazio01}: they  
exhibit a temperature $T_{BCS}$ at which the metallic grains  
placed on the sites of the array become (separately) superconducting  
and the Cooper pairs may be described by macroscopic wavefunctions.  
At a temperature $T_{BKT}<T_{BCS}$,  
the array undergo a BKT transition and the system - as a whole -  
becomes superconducting. 
 
\section{Observable Signature of the BKT Transition} 
 
In this Section we shall evidence that the 
experimental signature of the BKT transition in bosonic planar lattices  
is obtained by measuring, as a function of the temperature,  
the central peak of the interference pattern  
obtained after turning off the confining potentials.  

Firstly, we show that the peak of the momentum distribution at  
$\vec{k}=0$ is the direct analog of the magnetization of a  
finite size $2D$ $XY$ magnet. In fact, 
for the $XY$ magnets, the spins can be written  
as $\vec{S}_{\vec{j}}=(\cos{\theta_{\vec{j}}},\sin{\theta_{\vec{j}}})$ and  
the magnetization is defined as $M=(1/N) \cdot   
\langle \, \mid \sum_{\vec{j}} \vec{S}_{\vec{j}} \mid \, \rangle$  
where $\langle \cdots \rangle$ stands for the thermal average; 
a spin-wave analysis at low temperatures yields  
$M=(2N)^{-k_B T/8 \pi J}$ \cite{bramwell93,berezinskii73}.  
With discrete BECs at $T=0$, all the phases  
$\theta_{\vec{j}}$ are equal at the equilibrium and  
the lattice Fourier transform of $\psi_j$, 
\begin{equation}
\tilde{\psi}_{\vec{k}}=\frac{1}{N}  
\sum_{\vec{j}} \psi_{\vec{j}} \, \, \, 
e^{-i \vec{k} \cdot {\vec{j}}},
\label{FT}
\end{equation}  
exhibits a peak at $\vec{k}=0$ ($\vec{k}$  
is in the first Brillouin zone of the $2D$ square lattice); 
the magnetization is then  
\begin{equation} 
M=\langle \, \mid \tilde{\psi}_0 \mid \, \rangle. 
\label{M} 
\end{equation}    
The intuitive picture of the BKT transition is the following:  
at $T=0$, all the spins point in the same direction. 
As one can see from Fig.2(A), a single free vortex modifies the phase  
distribution far away from the vortex core and, thus, the modulus 
square of  
its lattice Fourier transform $\mid \tilde{\psi}_{\vec{k}} \mid^2$ 
has a minimum at $\vec{k}=0$. At variance, a vortex-antivortex   
pair [see Fig.2(B)] modifies the phase distribution only near the center of  
the pair (in this sense is a local {\em defect}) and 
its lattice Fourier transform  
has a maximum at $\vec{k}=0$. Analytical expressions for the  
lattice Fourier transform $\tilde{\psi}_{\vec{k}}$ of a single vortex and a  
vortex-antivortex pair may be worked out in detail 
and are reported in the Appendix B.  

Upon increasing the temperature, vortices are thermally induced.  
For $T < T_{BKT}$ only bound vortex pairs are present, and  
on average the spins continue to point in the same direction.  
When the condensates expand, a large peak (i.e., a magnetization)  
is observed in the central $\vec{k}=0$ momentum component,  
as shown in Fig.2(C). Rising further the temperature,  
due to the increasing number of vortex pairs, the central peak density  
decreases. For $T \approx T_{BKT}$,  
the pairs start to unbind and free vortices begin to appear  
[see Fig.2(D)], determining  
a sharp decrease around $T_{BKT}$ of the magnetization.  
At high temperatures, only free vortices are present, 
leading to a randomization  
of the phases and to a vanishing magnetization.  
In Fig.3 we plot the intensity of the central peak of the  
momentum distribution (normalized to the value at $T=0$) in a 2D  
lattice as a function of the reduced temperature $k_B T/J$,  
evidencing the sharp decrease  
of the magnetization around the BKT critical temperature.  
 
 
Let us now turn our attention to the interference patterns  
of the expanding condensates: after the bosonic system reaches the  
equilibrium, one may switch off both the harmonic trap and the optical 
lattice. At this time ($t=0$) the momentum distribution  
$\tilde{\psi}(\vec{p},t=0)$ is given by the Fourier transform of  
$\psi(\vec{r},t=0)=\sum_{\vec{j}} \psi_{\vec{j}} \Phi_{\vec{j}}(\vec{r})$: 
one finds 
\begin{equation}
\tilde{\psi}(\vec{p},t=0)= \tilde{\Phi}(\vec{p}) \tilde{\psi}_{\vec{k}},
\label{relation}
\end{equation}
where $\tilde{\Phi}(\vec{p})$ is the Fourier transform  
of the $3D$ Wannier functions and $\hbar \vec{k}=(p_x,p_y)$ is the  
momentum projection on the  
first Brillouin zone of the $2D$ optical lattice.  
Using a gaussian for the Wannier functions,  
$\Phi(\vec{r}) \propto e^{-x^2/2\sigma_x^2}  
e^{-y^2/2\sigma_y^2}e^{-z^2/2 \sigma_z^2}$  
(with $\sigma_z \gg \sigma \equiv \sigma_x=\sigma_y$,  
since $\tilde{\omega}_r \gg \omega_z$),  
one has $\tilde{\Phi}(\vec{p})=\chi(p_x) \chi(p_y) \chi(p_z) \propto  
e^{-\sigma^2 (p_x^2+p_y^2)/2\hbar^2}$, where $\chi(p_x) \propto  
e^{-\sigma_x^2 p_x^2/ 2 \hbar^2}$ and similarly for $\chi(p_y)$ and  
$\chi(p_z)$.  
Since for $s \gg1$ $\sigma / d =1/ \pi s^{1/4}$,  
the momentum distribution exhibits well  
pronounced peaks in the centers of each Brillouin zone: 
these peaks have different heights and the central $\vec{k}=0$ peak  
is the largest (see Figs.2 and 3 of \cite{greiner01}). 
  
At $t=0$ the amplitude of the $\vec{k}=0$ peak of the  
momentum distribution is simply given by the thermal average  
of $\tilde{\psi}_{0}$. By measuring the $\vec{k}=0$  
peak (i.e., $\langle \, \mid \tilde{\psi}_0  \mid^2 \,    \rangle$)  
at different temperatures, one obtains the results plotted in Fig.2.  
The figure has been obtained using a Monte Carlo simulation of the  
$XY$ magnet for a $40 \times 40$ array: we find $k_B T_C \approx 1.07 J$.  
In Fig.2 we also plot the low-temperature  
spin wave prediction \cite{berezinskii73} (solid line),  
as well as a fit, first used in  
\cite{bramwell93}, valid near $T_C$  
and derived form the renormalization group equations (dotted line).  
At times different from   
$t=0$, the density profiles  
are well reproduced by the free expansion of the ideal gas:  
one obtains $\tilde{\psi}(\vec{p},t)=\chi(p_z)  
e^{-i[(p_z+mgt)^3-p_z^3]/6m^2g\hbar} \tilde{\varphi}(p_x,p_y,t)$, i.e.  
a uniformly accelerating motion along $z$ and a free motion in the  
$x-y$ plane, with $\tilde{\varphi}(p_x,p_y,t)=  
\chi(p_x) \chi(p_y) \tilde{\psi}_{\vec{k}}e^{-i (p_x^2+p_y^2) t/2\hbar m}$ 
giving the central and lateral peaks of  
the momentum distribution as a function of time for different  
temperatures.  
 
An intense experimental work is now focusing on the  
Bose-Einstein condensation in two dimensions:  
at present a crossover to two-dimensional behaviour  
has been observed for $Na$ \cite{gorlitz01} and $Cs$ atoms \cite{hammes03}.  
Our analysis relies on two basic assumptions; namely,  
the validity of the tight-binding approximation  
for the Bose-Hubbard Hamiltonian and the requirement  
that the condensate in the optical lattice may be regarded as planar  
\cite{petrov00}. It is easy to see that the first assumption  
is satisfied if, at $T=0$, $V_0 \gg \mu$  
(where $\mu$ is the chemical potential), and, at finite temperature,  
$\hbar \tilde{\omega}_r \gtrsim k_B T$. The second assumption is much more  
restrictive since it requires freezing the transverse excitations;  
for this to happen one should require a condition on the transverse  
trapping frequency $\omega_z$. Namely, one should have that, at $T=0$,  
$\hbar \omega_z \gtrsim 8 K$ and that, at finite temperature,  
$\hbar \omega_z \gtrsim k_B T$ (since $\omega_z \ll \tilde{\omega}_r$,  
the latter condition also implies that $\hbar \tilde{\omega}_r  
\gtrsim k_B T$). In \cite{greiner01} it is   
$V_0=s \cdot k_B \cdot 0.15 \mu K$ and, for $s \gtrsim 20$,   
the tight-binding conditions are satisfied since  
$10 Hz \gtrsim 8K/2\pi \hbar$; furthermore, if $\omega_z=2 \pi \cdot 1kHz$,  
one may safely regard our finite temperature analysis to be valid at least  
up to $T \sim 50 nK$. We notice that the experimental signature for the   
BKT transition for a continuous  
(i.e., without optical lattice) weakly 
interacting 2D Bose gas \cite{prokofev01} is also given  
by the central peak of the atomic density of the  
expanding condensates. 
 
\section{Concluding Remarks} 
 
Our paper analyzes the finite-temperature phase transitions in $2D$  
lattices of BECs. Our study evidences the possibility that Bose-Einstein 
condensates loaded on a $2D$ optical lattice may exhibit  
- at finite temperature - a new coherent  
behaviour in which all the phases of the condensates 
located in each well  
of the lattice point in 
the same direction. The finite-temperature transition, which is due  
to the thermal atoms in each well, is mediated by vortex defects and may  
be experimentally detectable by looking at the 
interference patterns of the expanding condensates. Our analysis 
relies on the approximation that the effect of the shallow confining 
harmonic trap in the $x-y$ plane may be neglected: for a 
tight confinement, we expect that interesting mesoscopic effects 
will come to play, affecting the BKT transition.

We observe that 
for a $3D$ optical lattice (resulting form  
the potential $V_{opt}(\vec{r})=
V_0 [ \sin^2{(kx)} + \sin^2{(ky)} + \sin^2{(kz)}]$) 
with $N_0 \sim 100$, the system maps on the $3D$ $XY$  
model, which is the prototype of the universality class including the 
$\lambda$ normal-superfluid transition in $^4He$ \cite{tilley90}. 
Also in this case, the central peak of the interference pattern 
tends to zero at the critical temperature, Fig.4.  
We recall, however, that the $3D$ transition is a simple order-disorder  
transition (whose critical temperature $T_c$  
is given in the thermodynamic limit by $2.202J/k_B$  
\cite{gottlob93}), and is not mediated by  
the creation of vortices-antivortices pairs as in the 2D case. 
A natural question  
arising from the comparison of Figs.3 and 4 is how to 
clearly characterize the  
two transitions: to answer, one may observe that the critical exponent  
$\beta$ (defined by $M \propto (T_c-T)^\beta$) is 
$\beta \approx 0.23$ for the  
$2D$ $XY$ universality class \cite{bramwell93} and 
$\beta \approx 0.35$ for the  
$3D$ $XY$ universality class \cite{campostrini01}. 

In conclusion, our analysis strengthens 
- and extends at finite temperature - the  
striking and deep analogy of bosonic systems  
with superconducting Josephson junction arrays \cite{anderson98}.  
 
{\em Acknowledgements:} We thank Prof. E. Rastelli  
for stimulating discussions and for pointing our attention to Refs.  
\cite{bramwell93}. P.S. and A.T. enjoyed very much the  
residence at the CRS-BEC of Trento, where this work was initiated.   
This work has been supported by MIUR through grant No. 2001028294  
and by the DOE. 
 
\appendix 
 
\section{Variational Estimates of the Coefficients $K$ and $U$} 
 
In this Appendix we derive a variational 
estimate of the coefficients $K$ and $U$ entering the 
Bose-Hubbard Hamiltonian (\ref{BHM}).  
The tunneling rate $K$ and 
the coefficient of the nonlinear term $U$ are given by  
\begin{equation} 
K = - \int d\vec{r} \, \Bigg[ \frac{\hbar^2}{2m}  
\vec{\nabla} \Phi_{\vec{i}} \cdot \vec{\nabla} \Phi_{{\vec{j}}} +  
\Phi_{\vec{i}}  V_{ext} \Phi_{\vec{j}} \Bigg] 
\label{K} 
\end{equation} 
and  
\begin{equation} 
U = g_0 \int d\vec{r} \, \Phi_{\vec{j}}^4, 
\label{U} 
\end{equation} 
where $g_0=4 \pi \hbar^2 a / m$ and $\Phi_{\vec{j}}(\vec{r})$ is  
the Wannier wavefunction localized at the $\vec{j}$-th well  
and normalized to $1$. 
 
For a $2D$ optical lattice in the $x-y$ plane one has 
$V_{ext}=V_{opt}+V_m$, where the optical potential is 
$V_{opt}=V_0 [ \sin^2{(kx)} + \sin^2{(ky)} ]$ and  
the magnetic potential is  
$V_m=\frac{m}{2} [ \omega_r^2 (x^2+y^2) + \omega_z^2 z^2 ]$. 
Typical values of the parameters relevant for experiments 
\cite{greiner01} are 
$\omega_r , \omega_z \sim 2 \pi \cdot 10-100 \, Hz$, $\lambda=852 \, nm $, 
for a total number of sites $\sim 3000$ and an average number of 
particles per site $\sim 170$. 
The condition  $\omega_z \gg \omega_r$ should be imposed in order 
to have a $2D$ system. To get a variational estimate  
for $K$ and $U$ we assume that the Wannier wavefunction localized 
at the $j$-th well is given by  
\begin{equation} 
\Phi_{\vec{j}}(\vec{r})=C e^{-(x-x_j)^2/2\sigma_x^2}  
e^{-(y-y_j)^2/2\sigma_y^2} e^{-(z-z_j)^2/2\sigma_z^2}.  
\label{variazionale_2D} 
\end{equation}  
In Eq.(\ref{variazionale_2D}) 
the parameters $\sigma_x$, $\sigma_y$, and $\sigma_z$  
have to be determined variationally determined, 
$C$ is a normalization constant given by  
$C=(\pi^{3/2} \sigma^2 \sigma_z)^{-1/2}$ and $(x_j,y_j,z_j)$ 
are the coordinates of the center of the $j$-th well. 
Due to the symmetry of the external potential  
one has to set $\sigma_x=\sigma_y \equiv \sigma$: furthermore, since  
$\tilde{\omega}_r \gg \omega_r,\omega_z$, one has $\sigma_z \gg \sigma$.  
To fix $\sigma$ and $\sigma_z$ one recalls that 
the Gross-Pitaevskii energy is \cite{pitaevskii03} 
\begin{equation} 
{\cal E}[\Psi]=\int d\vec{r} \, \Bigg[ \frac{\hbar^2}{2m}  
(\vec{\nabla} \Psi)^2 + V_{ext} \mid \Psi \mid^2 + 
\frac{g_0}{2} \mid \Psi \mid^4 \Bigg]. 
\label{gp_energy}
\end{equation} 
In the tight-binding approximation,  
$\Psi(\vec{r},t)=\sum_j \psi_j(t) \Phi_j(\vec{r})$,  
one finds that the energy contribution $E_i$ of the $i$-th well to the total 
energy (\ref{gp_energy}) is  
$E_i \approx  
\int d\vec{r} \, \big[ \mid \psi_i \mid^2 \frac{\hbar^2}{2m}  
(\vec{\nabla}\Phi_i)^2 + V_{ext} \mid \psi_i \mid^2 \Phi_i^2 + 
\frac{g_0}{2} \mid \psi_i \mid^4 \Phi_i^4 \big]$, and, 
since $\mid \psi_i \mid^2 \approx N_0$, $N_0$ being 
the average value of particle per site, one gets
\begin{equation} 
E_i \approx  \int d\vec{r} \, \Bigg[ N_0 \frac{\hbar^2}{2m}  
(\vec{\nabla}\Phi_i)^2 + N_0 V_{ext} \Phi_i^2 + 
\frac{g_0}{2} N_0^2 \Phi_i^4 \Bigg].
\label{energia_singola_buca} 
\end{equation} 
Substituting (\ref{variazionale_2D}) in (\ref{energia_singola_buca}),  
one gets the following approximate expression  
for the energy in the $i$-th well   
\begin{equation}
\frac{E_i}{N_0} \approx \frac{\hbar^2}{2 m}  
\Bigg(\frac{1}{2\sigma_z^2} + \frac{1}{\sigma^2} \Bigg)+ 
\frac{m}{2} \Bigg(\omega_r^2 \sigma^2+\omega_z^2 \frac{\sigma_z^2}{2} \Bigg)+
V_0 k^2 \sigma^2 + \frac{g_0 N_0}{2(2\pi)^{3/2}\sigma^2\sigma_z}.
\label{e_i}
\end{equation} 
The optimal values of $\sigma$ and $\sigma_z$ are determined requiring that 
$\partial E_i / \partial \sigma=0$ and 
$\partial E_i / \partial \sigma_z=0$. 
It should be stressed that being $\sigma$ 
the variational width in the directions $x$ and $y$, 
it should be less than $\lambda/2$, the distance 
between minima of the periodic potential. 
We notice also that for realistic values of the parameters, 
the variational width $\sigma$ in the $x$ and $y$ directions 
has a very weak dependence on the mean field term (and therefore 
to the number of particles in each site): in fact, 
putting $\sigma \sim 0.2 \mu m$  
and $\sigma_z \sim 5 \mu m$, one finds  
$\hbar^2/2m\sigma^2 \sim 10^{-23} erg 
\gg g_0 N_0 / 2(2\pi)^{3/2}\sigma^2\sigma_z  
\sim 10^{-29} erg$, one sees that the last term in Eq.(\ref{e_i}) 
contributes very little to the determination of $\sigma$. 
Since $V_0 k^2 \gg m \omega_r^2/2$, 
from Eq.(\ref{e_i}) one then finds that the dependence of $E_i$ on 
$\sigma$ is given by $E_i(\sigma) \approx N_0 
(\hbar^2/2m\sigma^2 +V_0 k^2 \sigma^2)$: from 
$\partial E_i / \partial \sigma=0$ one obtains 
$\sigma^4=\hbar^2/2m k^2 V_0$, which can be cast in the form 
$\sigma=\lambda \, \chi$, with
\begin{equation}
\chi=\frac{1}{2\pi s^{1/4}}
\label{chi}
\end{equation}
and $s=V_0/E_R$. 

We are now in position to have an estimate for $K$ and $U$: 
by using the ansatz (\ref{variazionale_2D}) in 
Eqs.(\ref{K})-(\ref{U}) we get
\begin{equation}
\frac{K}{E_R}=\frac{1}{4\pi^2 \chi^4} \Bigg( \frac{1}{16} - 
\chi^2 \Bigg) \, e^{-1/16 \chi^2}- s e^{-1/16 \chi^2}
\label{K_2D}
\end{equation}   
and
\begin{equation}
\frac{U}{E_R}=\frac{2 m g_0}{(2 \pi)^{7/2} \hbar^2 \sigma_z \chi^2}.
\label{U_2D}
\end{equation}  

The variational procedure used works well also for $1D$ and $3D$ 
optical lattices: in a $1D$ optical lattice in the $x$ direction one has 
$V_{ext}=V_{opt}+V_m$, where $V_{opt}=V_0 \sin^2{(kx)}$ and 
$V_m=\frac{m}{2} [ \omega_x^2 x^2 + \omega_\perp^2 (y^2+z^2) ]$. 
Typical values of the parameters relevant for experiments 
\cite{cataliotti01} are 
$\omega_x \approx 2 \pi \cdot 10 \, Hz$, 
$\omega_\perp \approx 2 \pi \cdot 100 \, Hz$, $\lambda=795 \, nm $, 
for a total number of sites $\sim 100$ and an average number of 
particles per site $\sim 1000$. 
The variational ansatz is still given by Eq.(\ref{variazionale_2D}), 
but now $\sigma_y=\sigma_z \equiv \sigma_\perp$ and 
$\sigma_x \equiv \sigma$ is the variational width 
in the laser direction. Proceeding as before,  
one obtains that $\sigma=\lambda \, \chi$, 
with $\chi$ still given by Eq.(\ref{chi}). 
The tunneling rate $K$ is given by  
\begin{equation}  
\frac{K}{E_R} = \frac{1}{4\pi^2 \chi^4} 
\Bigg(\frac{1}{16}-\frac{\chi^2}{2} \Bigg) e^{-1/16\chi^2} -
\frac{s}{2} \, (1+e^{-4\pi^2 \chi^2}) \, e^{-1/16\chi^2},
\label{K_1D}
\end{equation}
and the nonlinear coefficient $U$ reads $U/E_R=2 m \lambda g_0 / 
(2 \pi)^{7/2} \hbar^2 \sigma_\perp^2 \chi$ where $\chi$ is given 
by Eq.(\ref{chi}).

For a $3D$ optical lattice one has 
$V_{ext}=V_{opt}+V_m$, where $V_{opt}=V_0 
[ \sin^2{(kx)} + \sin^2{(ky)} + \sin^2{(kz)}]$ and 
$V_m=\frac{m}{2} \omega (x^2+y^2+z^2)$. 
Typical experimental values from \cite{greiner02} are 
$\omega \approx 2 \pi \cdot 50\, Hz$, 
$\lambda=852 \, nm $, 
for a total number of sites $\sim 10^5$ and an average number of 
particles per site $\sim 1$. 
The variational ansatz is still given by Eq.(\ref{variazionale_2D}), 
but now $\sigma_x=\sigma_y=\sigma_z \equiv \sigma$. 
One obtains that $\sigma=\lambda \, \chi$, 
with $\chi$ given as before by Eq.(\ref{chi}). 
The tunneling rate $K$ is given by  
\begin{equation}  
\frac{K}{E_R} = \frac{1}{4\pi^2 \chi^4} 
\Bigg(\frac{1}{16}-\frac{3 \chi^2}{2} \Bigg) e^{-1/16\chi^2} -
s \, \Bigg(\frac{3}{2}-\frac{1}{2}e^{-4\pi^2 \chi^2}\Bigg) \, e^{-1/16\chi^2},
\label{K_3D}
\end{equation}
and the nonlinear coefficient $U$ reads $U/E_R=2 m g_0 / 
(2 \pi)^{7/2} \hbar^2 \lambda \chi^3$.

We observe that the tunneling rates $K$ for $1D$, $2D$ and $3D$ 
optical lattices, given respectively by 
Eqs.(\ref{K_1D}), (\ref{K_2D}) and (\ref{K_3D}),  
can be compactly written as  
\begin{equation}  
\frac{K}{E_R} = \frac{1}{4\pi^2 \chi^4} 
\Bigg(\frac{1}{16}-\frac{(j+1) \chi^2}{2} \Bigg) e^{-1/16\chi^2} -
s \, \Bigg( \frac{j+1}{2}-\frac{j-1}{2}e^{-4\pi^2 \chi^2} 
\Bigg) \, e^{-1/16\chi^2},
\label{K_general}
\end{equation}
where $j=D-1$.

\section{Fourier Transform of a Vortex on a Lattice}

In this Appendix we report the analytical 
computation of the lattice Fourier transform $\tilde{\psi}_{\vec{k}}$ 
of single vortex on a lattice described by 
$\psi_{\vec{j}}=\exp{(i \theta_{\vec{j}})}$; 
namely, one should compute
\begin{equation}
\tilde{\psi}_{\vec{k}}=\frac{1}{N}  
\sum_{\vec{j}} \psi_{\vec{j}} \, \, \, 
e^{-i \vec{k} \cdot {\vec{j}}}=\frac{1}{N}  
\sum_{\vec{j}} e^{-i \vec{k} \cdot {\vec{j}}+i\theta_{\vec{j}}}.
\label{FT_app}
\end{equation}
In Eq.(\ref{FT_app}), $\vec{j}=(j_x,j_y)$ denotes the sites 
of a square lattice (having $N$ sites) 
and $\vec{k}=(k_x,k_y)$ the (quasi)momentum, 
with $k_x$ and $k_y$ valued between $-\pi$ and $\pi$. 
The relationship between the lattice Fourier transform 
(\ref{FT_app}) and the momentum distribution $\tilde{\psi}(\vec{p})$ 
in real space is provided by Eq.(\ref{relation}). 

To simplify matters, in this Appendix we set the origin $O$ 
of the coordinates in the center 
of the central plaquette: setting to $1$ the lattice length, the lattice 
sites are labeled by $\vec{j}=(j_1/2,j_2/2)$ 
with $j_1$ and $j_2$ positive and negative {\em odd} integer numbers. 

The phase distribution of a vortex is 
depicted in Fig.2(A) and is such that $\theta_{\vec{j}}$ equals 
the polar angle $\phi$ with respect to the origin $O$ (e.g., 
at the site $(1/2,1/2)$ $\theta_{\vec{j}}$ equals $\pi / 4$, 
at the site $(3/2,1/2)$ 
is $\arctan{(1/3)}$, and so on; an antivortex would be characterized 
by $-\phi$). For future convenience, we shall label the site 
$(j_1+1/2,j_2+1/2)$ with the index $\ell=(\mid j_1 \mid + \mid j_2 \mid)/2$: 
as a consequence, the four sites $(1/2,1/2)$, 
$(1/2,-1/2)$, $(-1/2,1/2)$, and $(1/2,1/2)$ of the central plaquette 
will be associated to $\ell=1$; the eight sites 
$(3/2,1/2)$, $(1/2,3/2)$, $\cdots$, $(-3/2,1/2)$ 
will be associated to $\ell=2$. 
In other words, the sites with index $\ell$ have chemical 
distance $\ell-1$ from the four sites of the central plaquette. 
Of course, the number 
of sites with index $\ell$ is $4 \ell$. If, for simplicity, 
one considers only sites with $\ell$ going from $1$ to a maximum value 
${\cal L}$, then the total number of sites is 
\begin{equation}
N({\cal L})=\sum_{\ell=1}^{\cal L} 4 \ell=2 {\cal L} ({\cal L}+1).
\label{totale}
\end{equation} 

To evaluate the lattice Fourier transform (\ref{FT_app}), 
one may conveniently separate the sum 
$\sum_{\vec{j}} e^{-i \vec{k} \cdot {\vec{j}}+i\theta_{\vec{j}}}$ in 
$\cal L$ sums over sites having the same index $\ell$; for instance, 
for the four sites $(1/2,1/2)$, 
$(1/2,-1/2)$, $(-1/2,1/2)$, and $(1/2,1/2)$ of the central plaquette, 
one gets
\begin{equation}
\sum_{\vec{j}(1)} 
e^{-i \vec{k} \cdot {\vec{j}}+i\theta_{\vec{j}}} = 
4 \Bigg[ \sin{\phi^{(1)}_1} \, \cos{\frac{k_x}{2}} \, \sin{\frac{k_y}{2}} 
- i \cos{\phi^{(1)}_1} \, \sin{\frac{k_x}{2}} \, \cos{\frac{k_y}{2}}   \Bigg]
\label{sum_elle_1}
\end{equation}
where $\phi^{(1)}_1=\pi/4$ and 
the sum is restricted only to the sites having $\ell=1$. 
A straightforward generalization of Eq.(\ref{sum_elle_1}) shows that for 
the sum on the 
$4 \ell$ sites having a fixed $\ell$ one has
$$
\tilde{\psi}_{\vec{k}}^{(\ell)} \equiv \sum_{\vec{j}(\ell)} 
e^{-i \vec{k} \cdot {\vec{j}}+i\theta_{\vec{j}}} = 
4 \sum_{m=1}^{\ell} 
\Bigg[ \sin{\phi_m^{(\ell)}} 
\cos{ \frac{(2n-2m+1)k_x}{2}} \, \sin{ \frac{(2m-1)k_y}{2}} + 
$$
\begin{equation}
- i \cos{\phi_m^{(\ell)}} \, 
\sin{ \frac{(2n-2m+1)k_x}{2}} \, \cos{ \frac{(2m-1)k_y}{2}} \Bigg]
\label{sum_elle}
\end{equation}
where 
\begin{equation}
\phi_m^{(\ell)}=\arctan{\Bigg[ \frac{2m-1}{2n-2m+1} \Bigg]}
\label{theta_m_n}
\end{equation}
is an angle taking values between $0$ and $\pi / 2$. 

By using Eq.(\ref{sum_elle}), the lattice Fourier transform (\ref{FT_app}) 
may be compactly written as
\begin{equation}
\tilde{\psi}_{\vec{k}}=\frac{1}{N({\cal L})}  
\sum_{\ell=1}^{\cal L} \tilde{\psi}_{\vec{k}}^{(\ell)}.
\label{FT_an}
\end{equation}
     
If in Eq.(\ref{FT_an}) one puts $k_y=0$, one gets 
\begin{equation}
\tilde{\psi}_{\vec{k}}=\frac{-4i}{N({\cal L})}  
\sum_{\ell=1}^{\cal L} \sum_{m=1}^{\ell} \cos{\phi^{(\ell)}_m} 
\sin{\frac{(2n-2m+1)k_x}{2}};
\label{FT_an_simpl1}
\end{equation}
by introducing the index $\tilde{m}=\ell-m$ and rearranging conveniently 
the partial sums in Eq.(\ref{FT_an_simpl1}) one obtains 
\begin{equation}
\tilde{\psi}_{\vec{k}}=
\sum_{\tilde{m}=0}^{{\cal L}-1} a_{\tilde{m}} 
\sin{\frac{(2\tilde{m}+1)k_x}{2}}
\label{FT_an_simpl2}
\end{equation}
with
$$
a_{\tilde{m}}=\frac{-4i}{N({\cal L})} \sum_{n=\tilde{m}+1}^{\cal L} 
\frac{2 \tilde{m}+1}{\sqrt{(2 \tilde{m}+1)^2+(2n-2 \tilde{m}-1)^2}}.
$$
Eq.(\ref{FT_an_simpl2}) clearly shows that for $k_x=0$ one has 
$\tilde{\psi}_0=0$: of course, with a single vortex, the magnetization 
(\ref{M}) is zero; furthermore, in the thermodynamic limit ${\cal L} \to 
\infty$, $a_{\tilde{m}} \to 0$, in agreement the Mermin-Wagner theorem. We
observe Eq.(\ref{FT_an}) may be easily applied to the analysis 
of the lattice Fourier transform of a vortex-antivortex pair.

\begin{figure} 
\centerline{\psfig{figure=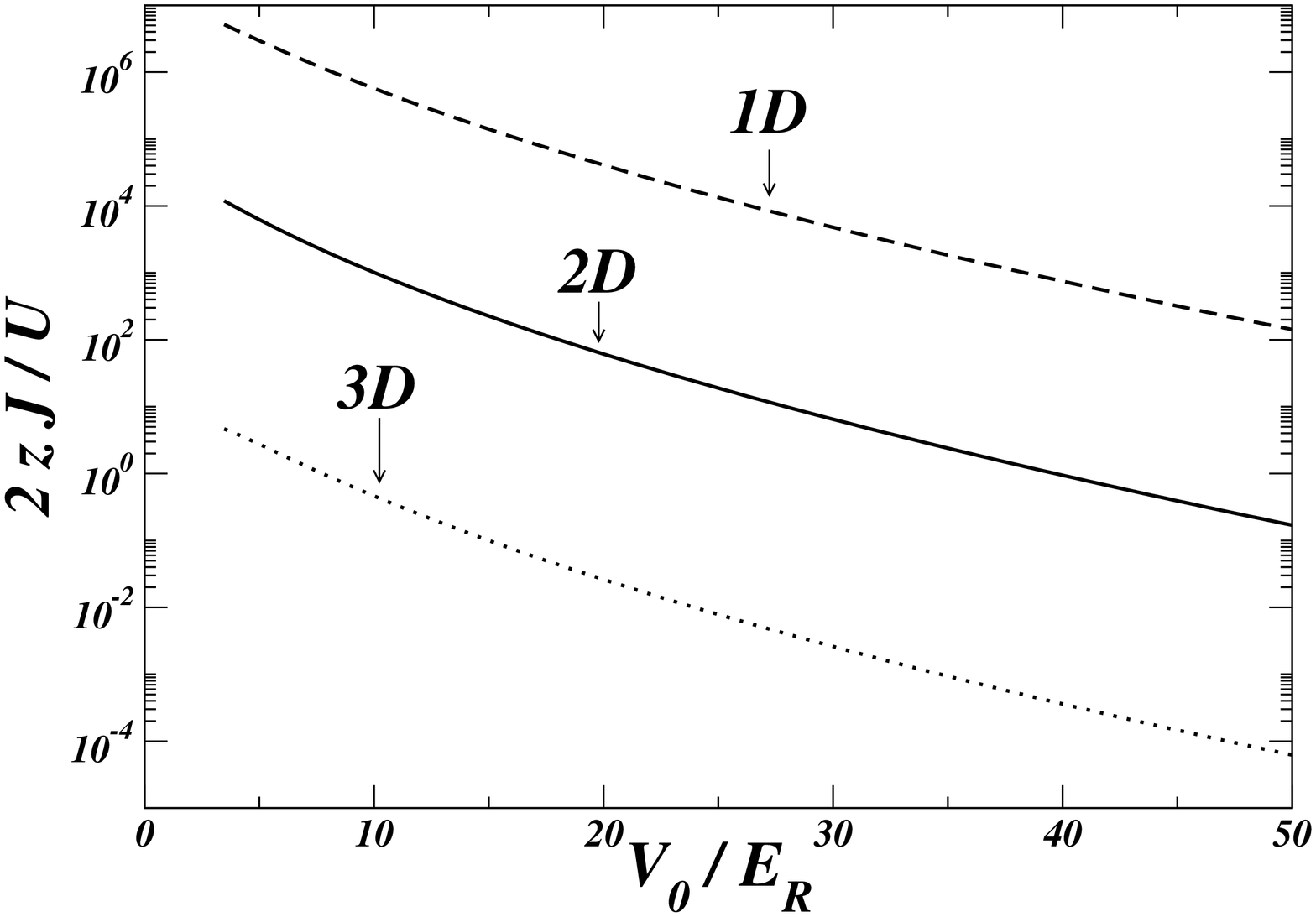,width=60mm,angle=270}} 
\caption{$J/U$ as a function of $V_0/E_R$ for $1D$, $2D$ and  
$3D$ optical lattices ($z$ is the number of nearest neighbours,  
which is respectively $2$, $4$, and $6$).  
The experimental values are taken, respectively, from 
\protect\cite{cataliotti01}, \protect\cite{greiner01},  
and \protect\cite{greiner02}, with an average number of $^{87} Rb$ atoms  
$N_0=1000$ ($1D$), $170$ ($2D$) and $1$ ($3D$). 
} 
\label{fig1} 
\end{figure} 
 
\begin{figure} 
\centerline{\psfig{figure=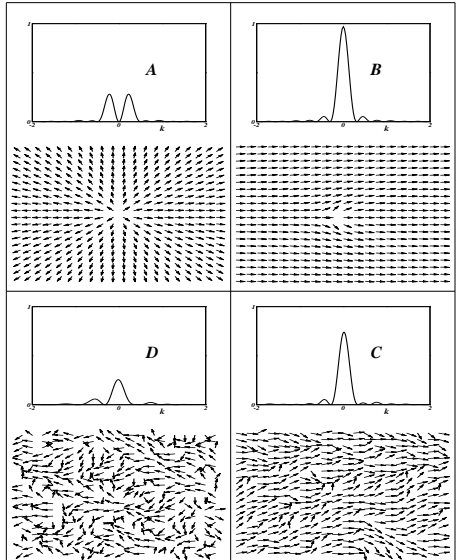,width=60mm,angle=0}} 
\caption{Lattice Fourier transform $\mid \tilde{\psi}_{\vec{k}} \mid^2$ 
(for $k_y=0$) with:  
(A) a single lattice vortex; (B) a lattice vortex-antivortex pair;  
(C) $T=0.5 J/k_B$; (D) $T = 1.1 J/k_B$.  
Figures (C) and (D) are Monte Carlo snapshots after reaching equilibrium.  
The lattice is $20 \times 20$.}  
\label{fig2} 
\end{figure} 
 
\begin{figure} 
\centerline{\psfig{figure=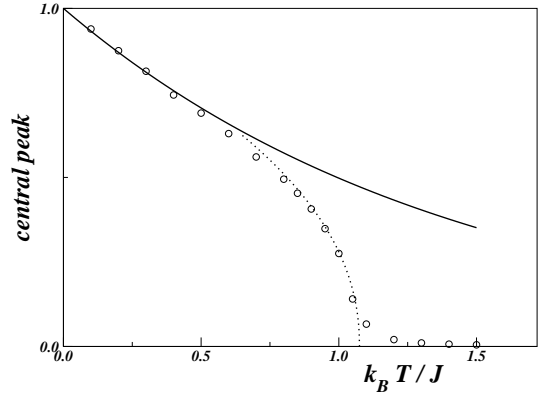,width=60mm,angle=270}} 
\caption{Intensity of the central peak of the momentum distribution  
(normalized to the value at $T=0$) as a function of the reduced temperature  
$t=k_B T/J$ in a 2D lattice. Empty circles: Monte Carlo simulations;  
solid line: low-temperature spin wave prediction;  
dotted line: fit near $T_C \approx 1.07  
J/k_B$ as in \protect\cite{bramwell93} (in this figure $L=40$).}  
\label{fig3} 
\end{figure} 

\begin{figure} 
\centerline{\psfig{figure=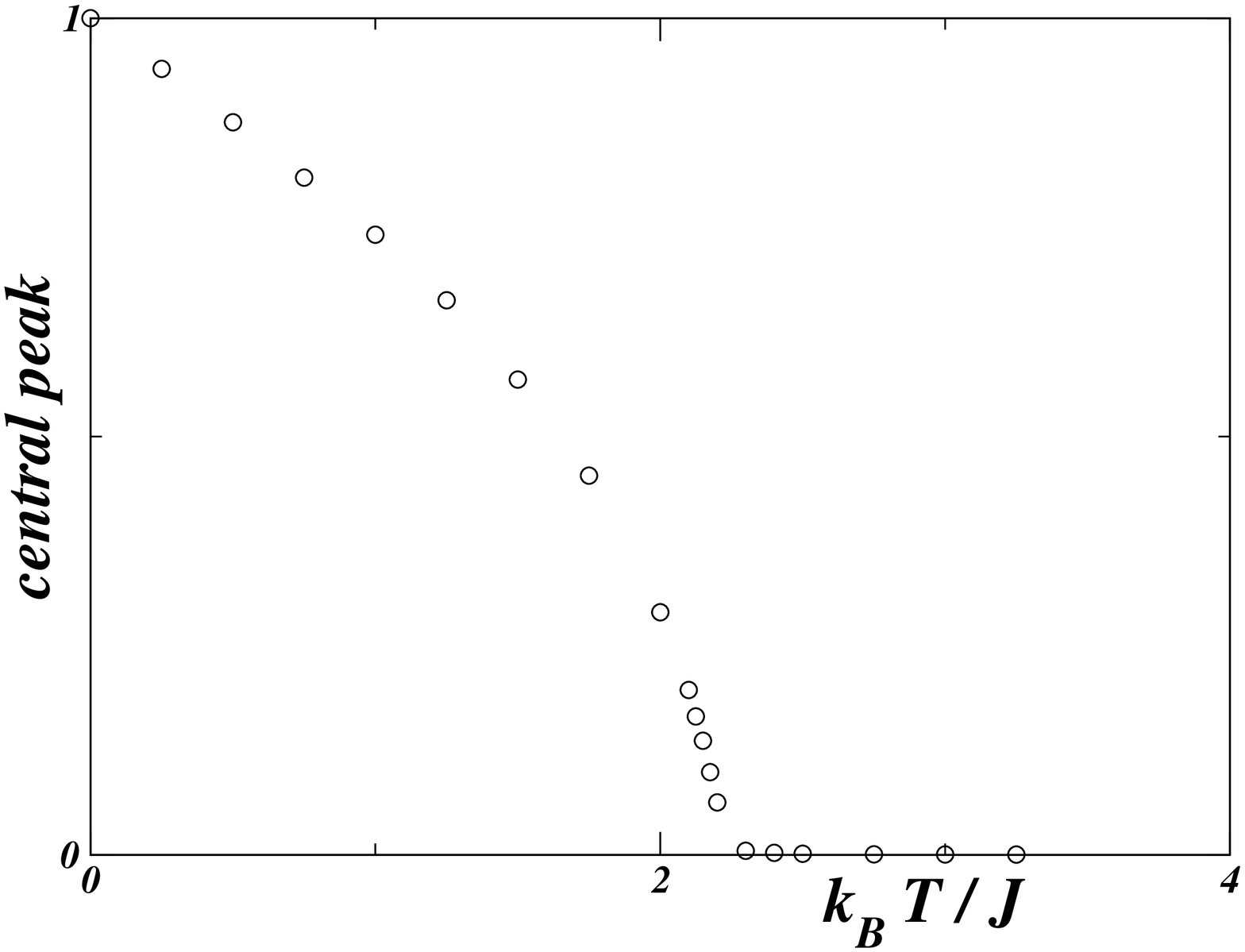,width=60mm,angle=270}} 
\caption{Monte Carlo results for the intensity 
of the central peak of the momentum distribution  
(normalized to the value at $T=0$) as a function of the reduced 
temperature  $t=k_B T/J$ in a 3D lattice (with $20^3$ sites).}  
\label{fig4} 
\end{figure}  
 

\begin{thebibliography}{10} 

\bibitem{anderson98} B. P. Anderson and M. A. Kasevich, 
Science {\bf 282}, 1686 (1998). 
 
\bibitem{cataliotti01} F. S. Cataliotti, S. Burger, 
C. Fort, P. Maddaloni, F. Minardi, A. Trombettoni, 
A. Smerzi, and M. Inguscio, Science {\bf 293}, 843 (2001).  
 
\bibitem{morsch01} O. Morsch, J. H. M\"uller, 
M. Cristiani, D. Ciampini, and E. Arimondo   
Phys. Rev. Lett. {\bf 87}, 140402 (2001).  
 
\bibitem{hensinger01} W. K. Hensinger {\em et al.},  
Nature {\bf 412}, 52 (2001).  
 
\bibitem{eiermann03} 
B. Eiermann, P. Treutlein, T. Anker, 
M. Albiez, M. Taglieber, K.-P. Marzlin, and M. K. Oberthaler 
Phys. Rev. Lett. {\bf 91}, 060402 (2003). 
 
\bibitem{greiner01}  
M. Greiner, I. Bloch, O. Mandel, T. W. H\"ansch, and 
T. Esslinger,  Phys. Rev. Lett. {\bf 87}, 160405 (2001). 
 
\bibitem{greiner02} M. Greiner, O. Mandel, T. Esslinger, 
T. W. H\"ansch, and I. Bloch, 
Nature {\bf 415}, 39 (2002). 
 
\bibitem{beasley79} M. R. Beasley, J. E. Mooij, and T. P. Orlando,  
Phys. Rev. Lett. {\bf 42}, 1165 (1979). 
 
\bibitem{resnick81} D. J. Resnick {\em et al.},  
Phys. Rev. Lett. {\bf 47}, 1542 (1981). 
 
\bibitem{nelson83} D. R. Nelson, in {\em Phase Transitions and  
Critical Phenomena}, vol. 7, eds. C. Domb and J. L. Lebowitz  
(New York, Academic Press, 1983) and references therein. 
 
\bibitem{minnaghen87} P. Minnaghen, Rev. Mod. Phys. {\bf 59},  
1001 (1987). 
 
\bibitem{kadanoff00} L. P. Kadanoff, {\em Statistical Physics: Statics,  
Dynamics and Renormalization} (Singapore, World Scientific, 2000),  
Chapt.s 16-17 and reprints therein. 
 
 
\bibitem{bishop78}  D. J. Bishop and J. D. Reppy,  
Phys. Rev. Lett. {\bf 40}, 1727 (1978). 

\bibitem{berezinskii73} V. L. Berezinskii and A. Ya. Blank, Sov. Phys. JETP  
{\bf 37}, 369 (1973). 

\bibitem{bramwell93} S. T. Bramwell and P. C. W. Holdsworth,  
J. Phys.: Condensed Matter {\bf 5}, L53 (1993);  
Phys. Rev. B {\bf 49}, 8811 (1994). 
 
\bibitem{pedri03} P. Pedri and L. Santos,  
Phys. Rev. Lett. {\bf 91}, 110401 (2003). 
 
\bibitem{jaksch98} D. Jaksch, C. Bruder, J. I. Cirac, 
C. W. Gardiner, and P. Zoller,
Phys. Rev. Lett. {\bf 81}, 3108 (1998). 
 
\bibitem{fisher89} M. P. A. Fisher, P. B. Weichman, G. Grinstein, 
and D. S. Fisher, Phys. Rev. B {\bf 40}, 546 (1989).  
 
 
 
 
\bibitem{simanek94} E. Sim\`anek, {\em Inhomogeneous Superconductors}, 
Oxford University Press, New York, 1994. 
 
\bibitem{fazio01} R. Fazio and H. van der Zant,  
Phys. Rep. {\bf 355}, 235 (2001). 

\bibitem{fisher88} M. P. A. Fisher and G. Grinstein, 
Phys. Rev. Lett. {\bf 60}, 208 (1988). 
 
\bibitem{kampf93} A. P. Kampf and G. T. Zimanyi,  
Phys. Rev. B {\bf 47}, 279 (1993). 
 
\bibitem{tsuchiya03} S. Tsuchiya and A. Griffin, cond-mat/0311321.  

\bibitem{gupta88} R. Gupta, J. DeLapp, G. G. Batrouni, 
G. C. Fox, C. F. Baillie, and J. Apostolakis, 
Phys. Rev. Lett. {\bf 61}, 1996 (1988).
 

\bibitem{smerzi04} A. Smerzi, P. Sodano, and A. Trombettoni,
J. Phys. B {\bf 37}, S265 (2004).
 
\bibitem{rojas96} C. Rojas and J. V. Jos\'e,  
Phys. Rev. B {\bf 54}, 12361 (1996). 

\bibitem{pitaevskii03} L. P. Pitaevskii and S. Stringari, 
{\it Bose-Einstein Condensation} 
(Oxford University Press, Oxford, 2003).
 
 
 
 
 
\bibitem{gorlitz01} A. G\"orlitz {\em et al.},  
Phys. Rev. Lett. {\bf 87}, 130402 (2001). 
 
\bibitem{hammes03} M. Hammes, D. Rychtarik, 
B. Engeser, H.-C. N\'agerl, and R. Grimm,  
Phys. Rev. Lett. {\bf 90}, 173001 (2001). 
 
\bibitem{petrov00} D. S. Petrov, M. Holzmann, and 
G. V. Shlyapnikov, Phys. Rev. Lett. {\bf 84}, 2551 (2000). 
 
 
\bibitem{prokofev01} N. Prokof\'ev, O. Ruebenacker, and B. Svistunov, 
Phys. Rev. Lett. {\bf 87}, 270402 (2001). 

\bibitem{tilley90} D. R Tilley and J. Tilley,  
{\em Superfluidity and Superconductivity} 
(Bristol, Adam Hilger, 1990). 
 
\bibitem{gottlob93} A. P. Gottlob and M. Hasenbusch,  
Physica A {\bf 201}, 593 (1993). 
 
\bibitem{campostrini01} M. Campostrini, M. Hasenbusch,  
A. Pelissetto, P. Rossi, and E. Vicari, Phys. Rev. {\bf B 63},  
214503 (2003). 
 
 
\end{thebibliography}
\end{document}